\documentclass[a4paper,10pt,eqno]{article}
\usepackage{theorem}
\usepackage[dvipdfmx]{graphicx}
\usepackage{latexsym,amssymb,amsfonts,amsmath}
\usepackage{cite}
\usepackage{color}
\usepackage{comment}

\newtheorem{thm}{Theorem}[section]

\newtheorem{pro}[thm]{Proposition}

\theoremheaderfont{\scshape}

\newcommand{\qed}{\hfill $\Box$}

\newcommand{\ket}[1]{|#1\rangle}
\newcommand{\bra}[1]{\langle#1|}

\title{{\Large {\bf Partition of graphs and quantum walk based search algorithms}}}
\author{
{\small Yusuke Ide}
{\scriptsize School of Knowledge Science, 
Japan Advanced Institute of Science and Technology,}\\
{\scriptsize 1-1 Asahidai, Nomi 923-1292, Japan}\\
{\scriptsize e-mail: ide@jaist.ac.jp}\\
}
\date{\today }
\begin{document}
\maketitle

\par\noindent
\begin{small}
\par\noindent
{\bf Abstract}
\newline 
In this paper, we show reduction methods for search algorithms on graphs using quantum walks. By using a graph partitioning method called equitable partition for the the given graph, we determine ``effective subspace'' for the search algorithm to reduce the size of the problem. We introduce the equitable partition for quantum walk based search algorithms and show how to determine ``effective subspace'' and reduced operator. 
\end{small}

\setcounter{equation}{0}
\section{Introduction}
In the last two decades, the study of the quantum walks (QWs) has been extensively developed by many researchers. QWs can be viewed as quantum counterparts of usual random walks (RWs) but they have several different features from that of RWs. For example, if we consider the discrete-time QW on the one-dimensional lattice case with parameters corresponding to the simple RW, the position $X_{n}$ of the walker at time $n$ follows the following limit theorem \cite{Konno2002, Konno2005}:
\begin{align*}
\frac{X_{n}}{n}\Rightarrow Z \quad (n\to \infty).
\end{align*}
Here the random variable $Z$ has the following probability density function:
\begin{align*}
f(x)=
\begin{cases}
\displaystyle
\frac{1}{\pi(1-x^{2})\sqrt{1-2x^{2}}}, &\text{if }-1/\sqrt{2}< x < 1/\sqrt{2},\\
0, &\text{otherwise}.
\end{cases}
\end{align*}
Note that $\Rightarrow$ stands for the weak convergence. Let $\mathbb{E}_{n}$ be the expectation with respect to the probability distribution of $X_{n}$ for each $n=0,1,\ldots $. We also consider the characteristic function $\phi _{n}(\xi)=\mathbb{E}_{n}[\exp(i\xi X_{n}/n)]$ of $X_{n}/n$ for each $\xi \in \mathbb{R}$ where $i$ (resp. $\mathbb{R}$) denotes the imaginary unit (resp. the set of real numbers). Then $X_{n}/n\Rightarrow Z$ is equivalent to $\lim _{n\to \infty}\phi _{n}(\xi) = \int _{-\infty}^{\infty}e^{i\xi x}f(x)dx$ for all $\xi \in \mathbb{R}$.

\begin{figure}[htb]
\begin{minipage}{0.5\linewidth}
  \begin{center}
  \includegraphics[width=0.89\linewidth]{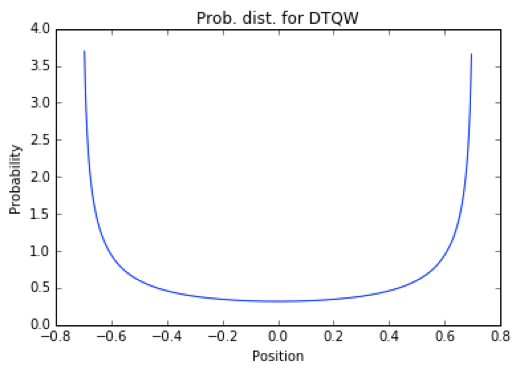}
  \caption{Limit distribution for discrete-time QWs.}
  \label{fig_Kdist}
  \end{center}
\end{minipage}
\begin{minipage}{0.5\linewidth}
  \begin{center}
  \includegraphics[width=0.9\linewidth]{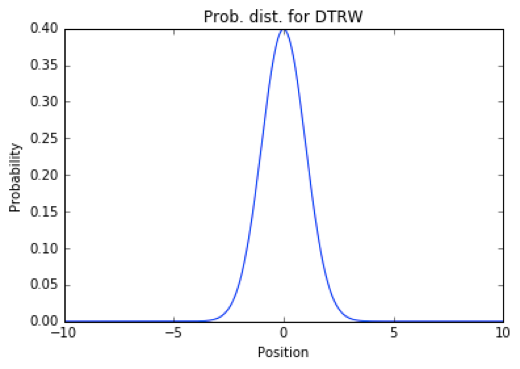}
  \caption{Limit distribution for RWs.}
  \label{fig_Gdist}
  \end{center}
\end{minipage}
\end{figure}

This theorem corresponds to the central limit theorem for the RWs. The theorem shows that the position of the quantum walker spreads in ballistic order (order $n$) not in diffusive order (order $\sqrt{n}$) of the random walker. The limit distribution (Fig.\ \ref{fig_Kdist}) is also different from the Gaussian distribution which is appear in the central limit theorem for the RWs  (Fig.\ \ref{fig_Gdist}). 

Recently, QWs have received much attention in various fields not only mathematical interests but also such as  experimental realization \cite{KarskiEtAl2009, SchreiberEtAl2012}, connection between topological phase \cite{AsbothEdge2015, KitagawaEtAl2010}, reduction method for radioactivity \cite{IchiharaEtAl2013, MatsuokaEtAl2011}, quantum search algorithms \cite{AmbainisEtAl2001, AmbainisEtAl2005, ChakrabortyEtAl2016, Childs2009, ChildsEtAl2003, ChildsGoldstone2004, LovettEtAl2010, ShenviEtAl2003}. There are good review articles for the theory of QWs such as \cite{Kempe2003, Kendon2007, Konno2008b, ManouchehriWang2013, Portugal2013, VAndraca2012}. 

In this paper, we show reduction methods for search algorithms on graphs using QWs. By using the equitable partition \cite{GodsilRoyle2001} for the graph, we determine ``effective subspace'' for the search algorithm to reduce the size of the problem. For this purpose, we review two types of QWs, discrete-time and continuous-time versions, in Sec.\ 2. In Sec.\ 3, we introduce QW based search algorithms. The main contribution of this paper is Sec.\ 4. In this section, we introduce the equitable partition for QW based search algorithms and show how to determine ``effective subspace'' and reduced operator. 

\section{Quantum walks}
\subsection{Discrete-Time Quantum Walk (DTQW)}
\begin{figure}[htb]
  \begin{center}
  \includegraphics{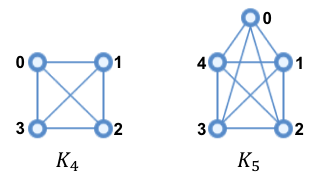}
  \caption{The complete graphs $K_{4}$ and $K_{5}$.}
  \label{fig_Kn}
  \end{center}
\end{figure}

Let $G=(V(G),E(G))$ be a simple graph (undirected graph without self-loops and multiple edges) with its vertex set $V(G)=\{0,1,\ldots, N-1\}$ and edge set $E(G)\subset V(G)\times V(G)$. As an example, for the complete graph on $N$ vertices $K_{N}$ which is the simple graph with $\binom{N}{2}$ edges (fully connected), the edge set is defined by $E(K_{N})=\{(j,k) : 0\leq j<k\leq N-1\}$. More concretely, for $K_{4}$ and $K_{5}$ (Fig.\ \ref{fig_Kn}), the edge sets are defined by $E(K_{4})=\{(0,1),(0,2),(0,3),(1,2),(1,3),(2,3)\}$ and $E(K_{5})=\{(0,1),(0,2),(0,3),(0,4),(1,2),(1,3),(1,4),\\(2,3),(2,4),(3,4)\}$.

\begin{figure}[htb]
  \begin{center}
  \includegraphics[width=0.225\linewidth]{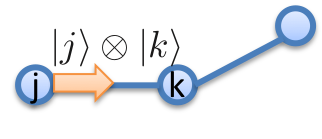}
  \caption{The state $\ket{j}\otimes \ket{k}$.}
  \label{fig_state}
  \end{center}
\end{figure}

Discrete-time quantum walk (DTQW) is a quantum dynamics on the graph with the following Hilbert space $\mathcal{H}_{DTQW}$:
\begin{align*}
\mathcal{H}_{DTQW}=\mathrm{Span}\left\{\ket{j}\otimes \ket{k}, \ket{k}\otimes \ket{j} : (j,k)\in E(G) \right\},
\end{align*}
where $\ket{j}= {}^{T}[0,\ldots ,0, \overbrace{1}^{\text{$j$-th}}, 0, \ldots ,0 ]$ is the $N$-dimensional standard basis (column vector, ${}^{T}A$ denotes the transpose of $A$) corresponding to the vertex $j$ and $\ket{j}\otimes \ket{k}$ represents the tensor product of the two bases $\ket{j}$ and $\ket{k}$. The Hilbert space $\mathcal{H}_{DTQW}$ is the Hilbert space spanned by the basis $\left\{\ket{j}\otimes \ket{k}, \ket{k}\otimes \ket{j} : (j,k)\in E(G) \right\}$. We usually call each element in the Hilbert space as state. The state $\ket{j}\otimes \ket{k}$ is interpreted as ``the state (direction) from the vertex $j$ to an adjacent vertex $k$'' (Fig.\ \ref{fig_state}).

\begin{figure}[htb]
  \begin{center}
  \includegraphics[width=0.5\linewidth]{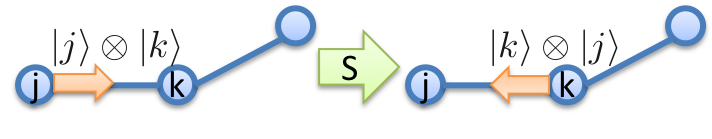}
  \caption{Flip-flop type shift.}
  \label{fig_shift}
  \end{center}
\end{figure}

\begin{figure}[htb]
  \begin{center}
  \includegraphics[width=0.55\linewidth]{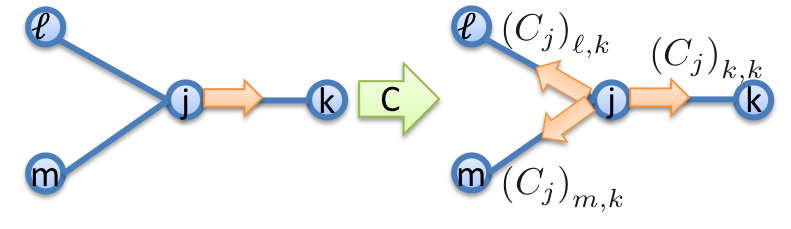}
  \caption{Action of the coin operator on the state $\ket{j}\otimes \ket{k}$.}
  \label{fig_coin}
  \end{center}
\end{figure}

There are many choices of the definition of the time evolution operator (unitary matrix) $U$ of DTQW on the graph $G$. Here we adopt the following definition:
\begin{align*}
U=SC.
\end{align*}
Where $S$ is called the flip-flop type shift operator which governs the motion of the walker. The definition of the shift is the following (Fig.\ \ref{fig_shift}):
\begin{align*}
S\left(\ket{j}\otimes \ket{k}\right)
=
\ket{k}\otimes \ket{j},\ \text{for $(j,k)\in E(G)$}.
\end{align*}
Also $C$ is called the coin operator which mixes the walker's states. The coin operator is given by
\begin{align*}
C
=
\sum_{j\in V(G)}\ket{j}\bra{j}\otimes C_{j},
\end{align*}
where $\bra{j}$ is the conjugate transpose of $\ket{j}$. The unitary matrix $C_{j}$ is a $d_{j}$ ($=$ the degree of the vertex $j$, i.e., the number of edges connected with the vertex $j$) -dimensional unitary matrix which is defined by
\begin{align*}
C_{j}\ket{k}
=
\sum_{(j,k^{\prime})\in E(G)}\left(C_{j}\right)_{k^{\prime},k}\ket{k^{\prime}},
\end{align*}
for each $k$ with $(j,k)\in E(G)$. From the definition of the coin operator $C$, we can see that 
\begin{align*}
C\ket{j}\otimes \ket{k}
=
\left(\sum_{j\in V(G)}\ket{j}\bra{j}\otimes C_{j}\right)\ket{j}\otimes \ket{k}
=
\ket{j}\otimes C_{j}\ket{k}
=
\sum_{(j,k^{\prime})\in E(G)}\left(C_{j}\right)_{k^{\prime},k}\ket{j}\otimes \ket{k^{\prime}},
\end{align*}
for each $(j,k)\in E(G)$. From this observation, we can say that the coin operator $C$ ``mixes'' each state corresponding to each vertex $j\in V(G)$ with suitable weights (Fig.\ \ref{fig_coin}).

There are many choices of $C_{j}$. One of a typical choice of $C_{j}$ is the following Grover's coin $U_{G}(j)$ which corresponds to the simple random walk on $G$:
\begin{align*}
U_{G}(j) = 2\ket{D_{j}}\bra{D_{j}}-I_{d_{j}},
\end{align*}
where $\ket{D_{j}}$ is called the diagonal sate corresponding to the vertex $j$ which is defined by
\begin{align*}
\ket{D_{j}}=\frac{1}{\sqrt{d_{j}}}\sum_{(j,k)\in E(G)}\ket{k},
\end{align*}
and $I_{k}$ denotes the identity matrix of order $k$.

For DTQW, we choose a unit vector (initial state) $\ket{\Psi_{N,0}}\in \mathcal{H}_{DTQW}$ of the walker. Then we consider the time evolution
\begin{align*}
\ket{\Psi_{N,n}} = U^{n}\ket{\Psi_{N,0}},
\end{align*}
for $n=0,1,\ldots $. Here $\ket{\Psi_{N,n}}$ is called the probability amplitude at time $n$. Using this probability amplitude, we define the probability
\begin{align*}
\mathbb{P}_{N,n}(s) = \lVert \left(\ket{s}\bra{s}\otimes I_{d_{s}}\right) \ket{\Psi_{N,n}} \rVert^{2},
\end{align*}
that the walker is found on $s\in V(G)$ at time $n$ ．

\subsection{Continuous-Time Quantum Walk (CTQW)}
\begin{figure}[htb]
\begin{minipage}{0.5\linewidth}
  \begin{center}
  \includegraphics[width=0.6\linewidth]{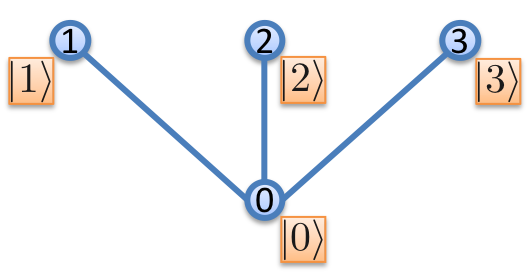}
  \caption{The basis of $\mathcal{H}_{CTQW}$.}
  \label{fig_basisCTQW}
  \end{center}
\end{minipage}
\begin{minipage}{0.5\linewidth}
  \begin{center}
  \includegraphics[width=0.25\linewidth]{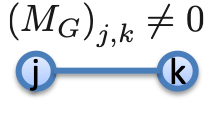}
  \caption{The weight of the edge $(j,k)\in E(G)$.}
  \label{fig_weightCTQW}
  \end{center}
\end{minipage}
\end{figure}

Again we consider a simple graph $G=(V(G),E(G))$ with $V(G)=\{0,1,\ldots, N-1\}$. Continuous-time quantum walk (CTQW) is a quantum dynamics on the graph with the following Hilbert space $\mathcal{H}_{CTQW}$ (Fig.\ \ref{fig_basisCTQW}):
\begin{align*}
\mathcal{H}_{CTQW}=\mathrm{Span}\left\{\ket{j} : j\in V(G) \right\}.
\end{align*}
On this Hilbert space $\mathcal{H}_{CTQW}$, we define the time evolution operator of CTQW. We choose an $N$-dimensional Hermitian matrix $M_{G}$ with $(M_{G})_{j,k}\neq 0$ when $(j,k)\in E(G)$ and $(M_{G})_{j,k}= 0$ when $(j,k)\notin E(G)$. The component $(M_{G})_{j,k}$ is interpreted as ``the weight of the edge $(j,k)\in E(G)$'' as Fig.\ \ref{fig_weightCTQW}. 

For the time $t\geq 0$, we define the time evolution operator 
\begin{align*}
U_{M_{G}}(t) = \exp(itM_{G}) = \sum_{k=0}^{\infty}\frac{(it)^{k}}{k!}M_{G}^{k},
\end{align*}
of CTQW on $G$ corresponding to $M_{G}$. We note that CTQW is nothing but quantum dynamics determined by the Schr\"odinger equation with its Hamiltonian $M_{G}$. We choose a unit vector (initial state) $\ket{\Psi_{N,0}}\in \mathcal{H}_{CTQW}$ then the time evolution
\begin{align*}
\ket{\Psi_{N,t}} = U_{M_{G}}(t)\ket{\Psi_{N,0}},
\end{align*}
for $t\geq 0$ is defined. Here $\ket{\Psi_{N,t}}$ represents the probability amplitude at time $t$. We define the probability that the walker is found on $s\in V(G)$ at time $t$ as
\begin{align*}
\mathbb{P}_{N,t}(s) = |\langle s | \Psi_{N,t}\rangle |^{2}.
\end{align*}
One of a typical choice of $M_{G}$ is the adjacency matrix
\begin{align*}
(A_{G})_{j,k} = 
\begin{cases}
1 & \text{if $(j,k)\in E(G)$},\\
0 & \text{otherwise}.
\end{cases}
\end{align*}
of the graph $G$. 

For example, we show the time evolution for $G=K_{2}$ the complete graph on $2$ vertices case. In this case, 
\begin{align*}
A_{K_{2}} = 
\begin{bmatrix}
0 & 1 \\
1 & 0
\end{bmatrix}. 
\end{align*}
Thus we have
\begin{align*}
A_{K_{2}}^{2k} = 
\begin{bmatrix}
1 & 0 \\
0 & 1
\end{bmatrix}
,\quad 
A_{K_{2}}^{2k+1} = 
\begin{bmatrix}
0 & 1 \\
1 & 0
\end{bmatrix}.
\end{align*}
Therefore we obtain
\begin{align*}
U_{A_{K_{2}}}(t)
&=
\exp(itA_{K_{2}}) = \sum_{k=0}^{\infty}\frac{(it)^{k}}{k!}A_{K_{2}}^{k}
\\
&=
\sum_{k=0}^{\infty}\frac{(-1)^{k}t^{2k}}{(2k)!}
\begin{bmatrix}
1 & 0 \\
0 & 1
\end{bmatrix}
+
i\sum_{k=0}^{\infty}\frac{(-1)^{k}t^{2k+1}}{(2k+1)!}
\begin{bmatrix}
0 & 1 \\
1 & 0
\end{bmatrix}
\\
&=
\begin{bmatrix}
\cos t & i \sin t \\
i \sin t & \cos t
\end{bmatrix}.
\end{align*}

If we consider the initial state as $\ket{\Psi_{n,0}}= \binom{1}{0}$ which is corresponding to the case that the walker starts from the vertex $0\in V(K_{2})$ then we can calculate 
\begin{align*}
\ket{\Psi_{N,t}}
&=
U_{A_{K_{2}}}(t)\ket{\Psi_{n,0}}
=
\begin{bmatrix}
\cos t & i \sin t \\
i \sin t & \cos t
\end{bmatrix}
\begin{bmatrix}
1\\
0
\end{bmatrix}
=
\begin{bmatrix}
\cos t \\
i \sin t 
\end{bmatrix}.
\end{align*}
Therefore we obtain
\begin{align*}
\mathbb{P}_{N,t}(0)
&=
|\langle 0\ket{\Psi_{N,t}}|^{2} = |\cos t|^{2} = \cos ^{2} t,
\\
\mathbb{P}_{N,t}(1)
&=
|\langle 1\ket{\Psi_{N,t}}|^{2} = |i \sin t|^{2} = \sin ^{2} t.
\end{align*}

\section{Quantum Walk based search algorithms}
\subsection{DTQW cases}
In this section, we consider discrete-time quantum walk based search (DTQW search) on simple graph $G=(V(G),E(G))$ with $V(G)=\{0,1,\ldots, N-1\}$. Let $w\in V(G)$ be the vertex which we want to find (marked vertex). For DTQW search, we use the following special coin operator:
\begin{align*}
C_{j}=
\begin{cases}
-I_{d_{w}}, &\text{if $j=w$,}\\
U_{G}(j), &\text{otherwise.}
\end{cases}
\end{align*}
Then we consider the DTQW on $G$ with the uniform initial state
\begin{align*}
|\Psi_{0}\rangle 
=
\frac{1}{\sqrt{2|E(G)|}}\sum_{j=0}^{N-1}\sum_{(j,k)\in E(G)}\ket{j}\otimes \ket{k}.
\end{align*}
The main task for DTQW search is finding a time $T=\mathcal{O}(\sqrt{N})$ such that we can attain
\begin{align*}
\mathbb{P}_{N,T}(w)
\approx 1\quad (N\to \infty).
\end{align*}

For example, for DTQW search on the hypercube case, there is a time $T=\mathcal{O}(\sqrt{N})$ such that the search is successful \cite{ShenviEtAl2003}. For the square lattice cases \cite{AmbainisEtAl2005}, there also exist times $T=\mathcal{O}(\sqrt{N})$.

\subsection{CTQW cases}
In this section, we consider continuous-time quantum walk based search (CTQW search) on simple graph $G=(V(G),E(G))$ with $V(G)=\{0,1,\ldots, N-1\}$. The Hermitian matrix $H_{G}$ for CTQW search on $G$ is defined by 
\begin{align*}
H_{G}
=
|w\rangle \langle w| + \gamma A_{G}\quad (\gamma \in \mathbb{R}),
\end{align*}
where $w\in V(G)$ is the marked vertex. Now we consider the CTQW on $G$ with its time evolution operator $U_{H_{G}}$. The main task for CTQW search is finding a suitable $\gamma \in \mathbb{R}$ and a time $T=\mathcal{O}(\sqrt{N})$ such that we can attain
\begin{align*}
\mathbb{P}_{N,T}(w)
\approx 1\quad (N\to \infty),
\end{align*}
with the uniform initial state
\begin{align*}
|\Psi_{0}\rangle 
=
\frac{1}{\sqrt{N}}\sum_{j=0}^{N-1}|j\rangle .
\end{align*}

For example, CTQW search on the complete graph, the hypercube and the square lattice cases, there exist suitable $\gamma \in \mathbb{R}$ and a time $T=\mathcal{O}(\sqrt{N})$ such that the search are successful \cite{ChildsGoldstone2004}. For the Erd\"os-R\'enyi random graph with suitable condition for the connection probability cases, there also exist suitable $\gamma \in \mathbb{R}$ and a time $T=\mathcal{O}(\sqrt{N})$ such that the search are successful for almost all generated graphs \cite{ChakrabortyEtAl2016}.

\section{Equitable partition and quantum walk search}
\subsection{Equitable partition of graphs for quantum walk search}
For both DTQW search and CTQW search, a partition of the graph so-called equitable partition \cite{GodsilRoyle2001} is very useful. Note that the equitable partition can be viewed as a generalization of the modular partition \cite{HabibPaul2010}. Without loss of generality,  we assume that the marked vertex is $0\in V(G)$. We consider the partition $(G_{\bar{0}}, G_{\bar{1}}, \ldots ,G_{\overline{J-1}})$ of $G=(V(G),E(G))$ which is satisfied the following $3$ conditions:
\begin{enumerate}
\item
$
V(G_{\bar{0}})=\{0\}.
$
\item
$
V(G)=\bigcup_{\bar{j}=0}^{J-1}V(G_{\bar{j}}),\ V(G_{\bar{j}})\cap V(G_{\bar{k}}) = \emptyset \ (\bar{j}\neq \bar{k}).
$
\item
For each $\bar{0}\leq \bar{j},\bar{k}\leq \overline{J-1}$, there exists a non-negative integer $d_{\bar{j},\bar{k}}$ such that $d_{\bar{j},\bar{k}}=| \{w\in V(G_{\bar{k}}) : (v,w)\in E(G)\}|$ for each $v\in V(G_{\bar{j}})$.
\end{enumerate}
This partition is a special case of so-called equitable partition. We use the notation $G_{\bar{j}}\sim G_{\bar{k}}$ if $d_{\bar{j},\bar{k}}\neq 0$. 

\begin{figure}[htb]
  \centering
  \includegraphics[width=0.25\linewidth]{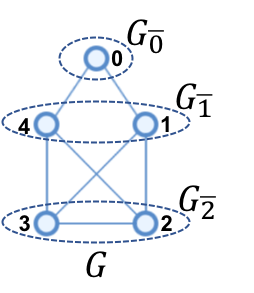}
  \caption{Partition of the graph $G$}
  \label{fig_partition}
\end{figure}

The partition $(G_{\bar{0}}, G_{\bar{1}}, \ldots ,G_{\overline{J-1}})$ of $G=(V(G),E(G))$ consists of $\bar{J}$ subgraphs. For each subgraph $G_{\bar{j}}$ is $d_{\bar{j},\bar{j}}$-regular graph. Especially, $G_{\bar{0}}$ is the null graph ($d_{\bar{0},\bar{0}}=0$) which consists of only the marked vertex $0\in V(G)$. In addition, for each vertex $v\in V(G_{\bar{j}})$ the number of edges connected to the vertices in $V(G_{\bar{k}})$ is $d_{\bar{j},\bar{k}}$ regardless the choice of vertex. In Fig.\ \ref{fig_partition}, we show an example of the partition. In this case, $V(G_{\bar{0}})=\{0\}, V(G_{\bar{1}})=\{1,4\}, V(G_{\bar{2}})=\{2,3\}$ and $d_{\bar{0},\bar{0}}=0, d_{\bar{0},\bar{1}}=2, d_{\bar{0},\bar{3}}=0, d_{\bar{1},\bar{0}}=1, d_{\bar{1},\bar{1}}=0, d_{\bar{1},\bar{2}}=2, d_{\bar{2},\bar{0}}=0, d_{\bar{2},\bar{1}}=2, d_{\bar{2},\bar{2}}=1$.
By using this partition, we can construct efficient eigenspace for both DTQW search and CTQW search. Note that for each graph $G$, the graph itself, i.e., $(G_{\bar{0}}, G_{\bar{1}}, \ldots ,G_{\overline{N-1}})$ with $V(G_{\bar{j}})=\{j\}$ for each $j\in V(G)$, is a trivial equitable partition. Therefore the proposed method works well for graphs with equitable partition consists of small numbers of subgraphs. 

\subsection{DTQW cases}
In this section, we consider DTQWs defined by ``Grover type coins'' on simple, connected graphs. For each $j\in V(G)$, we consider the following coin:
\begin{align*}
C_{j}
=
(\lambda_{j,1}-\lambda_{j,2})\ket{D_{j}}\bra{D_{j}}+\lambda_{j,2}I_{d_{j}},
\end{align*}
where $\lambda_{j,1}, \lambda_{j,2}\in \mathbb{C}$ with $|\lambda_{j,1}|=|\lambda_{j,2}|=1$. Note that if $\lambda_{j,1}\neq \lambda_{j,2}$ then this Grover type coin $C_{j}$ has two eiganvalues $\lambda_{j,1}$ and $\lambda_{j,2}$ with multiplicities $1$ and $N-1$, respectively. \

Here we consider the equitable partition $(G_{\bar{0}}, G_{\bar{1}}, \ldots ,G_{\overline{J-1}})$ with $n_{\bar{j}}=|V(G_{\bar{j}})|$ of $G$. We define the following uniform states related to the partition:
\begin{align*}
\ket{\bar{j}\bar{k}}
=
\begin{cases}
\displaystyle\frac{1}{\sqrt{n_{\bar{j}}d_{\bar{j},\bar{k}}}}\sum_{\substack{j\in V(G_{\bar{j}}), k\in V(G_{\bar{k}}),\\ (j,k)\in E(G)}}\ket{j}\otimes \ket{k}, &\text{if $G_{\bar{j}}\sim G_{\bar{k}}$},\\
\mathbf{0}, &\text{otherwise},
\end{cases}
\end{align*}
for each $\bar{0}\leq \bar{j},\bar{k}\leq \overline{J-1}$ where $\mathbf{0}$ is the zero vector of $\mathcal{H}_{DTQW}$. The state $\ket{\bar{j}\bar{k}}$ is the uniform state on the states corresponding to edges from $G_{\bar{j}}$ to $G_{\bar{k}}$. Assume that every vertices in $G_{\bar{j}}$ have the same coin $C_{\bar{j}}$ for $\bar{0}\leq \bar{j}\leq \overline{J-1}$ then we obtain
\begin{align*}
C\ket{\bar{j}\bar{k}}
&=
\frac{\lambda_{\bar{j},1}-\lambda_{\bar{j},2}}{d_{\bar{j}}}\sqrt{\frac{d_{\bar{j},\bar{k}}}{n_{\bar{j}}}}\left(\sum_{G_{\bar{j}}\sim G_{\bar{k^{\prime}}}}\sqrt{n_{\bar{j}}d_{\bar{j},\bar{k^{\prime}}}}\ket{\bar{j}\bar{k^{\prime}}}\right)
+
\lambda_{\bar{j},2}\ket{\bar{j}\bar{k}}
\\
&=
\frac{\lambda_{\bar{j},1}-\lambda_{\bar{j},2}}{d_{\bar{j}}}\left(\sum_{G_{\bar{j}}\sim G_{\bar{k^{\prime}}}}\sqrt{d_{\bar{j},\bar{k}}d_{\bar{j},\bar{k^{\prime}}}}\ket{\bar{j}\bar{k^{\prime}}}\right)
+
\lambda_{\bar{j},2}\ket{\bar{j}\bar{k}},
\end{align*}
with convention $G_{\bar{j}}\sim G_{\bar{j}}$ for $\bar{0}\leq \bar{j}\leq \overline{J-1}$. On the other hand, 
\begin{align*}
S\ket{\bar{j}\bar{k}}
&=
S\left(\frac{1}{\sqrt{n_{\bar{j}}d_{\bar{j},\bar{k}}}}\sum_{\substack{j\in V(G_{\bar{j}}), k\in V(G_{\bar{k}}),\\ (j,k)\in E(G)}}\ket{j}\otimes \ket{k}\right)
=
\frac{\sqrt{n_{\bar{k}}d_{\bar{k},\bar{j}}}}{\sqrt{n_{\bar{j}}d_{\bar{j},\bar{k}}}}\frac{1}{\sqrt{n_{\bar{k}}d_{\bar{k},\bar{j}}}}\sum_{\substack{k\in V(G_{\bar{k}}), j\in V(G_{\bar{j}}),\\ (k,j)\in E(G)}}\ket{k}\otimes \ket{j}
\\
&=
\ket{\bar{k}\bar{j}},
\end{align*}
where we use the relations $(j,k)\in E(G)$ then $(k,j)\in E(G)$ and $n_{\bar{j}}d_{\bar{j},\bar{k}}=n_{\bar{k}}d_{\bar{k},\bar{j}}$ coming from simplicity of the graph. This shows that the action of $U=SC$ is 
\begin{align*}
U\ket{\bar{j}\bar{k}}
&=
\frac{\lambda_{\bar{j},1}-\lambda_{\bar{j},2}}{d_{\bar{j}}}\left(\sum_{G_{\bar{j}}\sim G_{\bar{k^{\prime}}}}\sqrt{d_{\bar{j},\bar{k}}d_{\bar{j},\bar{k^{\prime}}}}\ket{\bar{k^{\prime}}\bar{j}}\right)
+
\lambda_{\bar{j},2}\ket{\bar{k}\bar{j}},
\end{align*}
for $\bar{0}\leq \bar{j},\bar{k}\leq \overline{J-1}$.

From this observation, the action of the time evolution operator $U=SC$ is closed on the subspace $\mathrm{Span}\{\ket{\bar{j}\bar{k}} : \bar{0}\leq \bar{j},\bar{k}\leq \overline{J-1}\}$ of $\mathcal{H}_{DTQW}$. Also the uniform initial state $\ket{\Psi_{0}}$ is represented by linear combination of uniform states related to the partition, 
\begin{align*}
|\Psi_{0}\rangle 
=
\frac{1}{\sqrt{2|E(G)|}}\sum_{j=0}^{N-1}\sum_{(j,k)\in E(G)}\ket{j}\otimes \ket{k}
=
\frac{1}{\sqrt{2|E(G)|}}\sum_{\bar{0}\leq \bar{j},\bar{k}\leq \overline{J-1}}\sqrt{n_{\bar{j}}d_{\bar{j},\bar{k}}}\ket{\bar{j}\bar{k}}.
\end{align*}
This shows that $\ket{\Psi_{0}}\in \mathrm{Span}\{\ket{\bar{j}\bar{k}} : \bar{0}\leq \bar{j},\bar{k}\leq \overline{J-1}\}$. We summarize this fact as the following proposition:
\begin{pro}\label{pro_DTQW}
Consider the DTQW defined by Grover type coin with the equitable partition $(G_{\bar{0}}, G_{\bar{1}}, \ldots , G_{\overline{J-1}})$ starting from the uniform initial state $\ket{\Psi_{0}}$. Then we have $\ket{\Psi_{0}}\in \mathrm{Span}\{\ket{\bar{j}\bar{k}} : \bar{0}\leq \bar{j},\bar{k}\leq \overline{J-1}\}$. Assume that every vertices in $G_{\bar{j}}$ have the same coin $C_{\bar{j}}$. Then the action of the time evolution operator $U=SC$ is closed on the subspace $\mathrm{Span}\{\ket{\bar{j}\bar{k}} : \bar{0}\leq \bar{j},\bar{k}\leq \overline{J-1}\}$ of $\mathcal{H}_{DTQW}$ as follows:
\begin{align}\label{eq_DTQW}
U\ket{\bar{j}\bar{k}}
&=
\frac{\lambda_{\bar{j},1}-\lambda_{\bar{j},2}}{d_{\bar{j}}}\left(\sum_{G_{\bar{j}}\sim G_{\bar{k^{\prime}}}}\sqrt{d_{\bar{j},\bar{k}}d_{\bar{j},\bar{k^{\prime}}}}\ket{\bar{k^{\prime}}\bar{j}}\right)
+
\lambda_{\bar{j},2}\ket{\bar{k}\bar{j}},
\end{align}
for $\bar{0}\leq \bar{j},\bar{k}\leq \overline{J-1}$.
\end{pro}

Proposition \ref{pro_DTQW} states a reduction method for the DTQW defined by Grover type coin with the equitable partition $(G_{\bar{0}}, G_{\bar{1}}, \ldots , G_{\overline{J-1}})$ starting from the uniform initial state $\ket{\Psi_{0}}$. Whenever we start from the uniform initial state $\ket{\Psi_{0}}$, we can only consider the action of the time evolution operator $U=SC$ on $\mathrm{Span}\{\ket{\bar{j}\bar{k}} : \bar{0}\leq \bar{j},\bar{k}\leq \overline{J-1}\}$ that is defined as Eq.\ \eqref{eq_DTQW}. This means that we can only consider at most $J^{2}$ dimensional subspace of $\mathcal{H}_{DTQW}$ which dimension is $2|E(G)|$. Applying Proposition \ref{pro_DTQW} to DTQW search, we have the following theorem:
\begin{thm}\label{thm_DTQW}
Consider the DTQW search on simple, connected graph $G$ with the equitable partition $(G_{\bar{0}}, G_{\bar{1}}, \ldots , G_{\overline{J-1}})$. Then we have $\ket{\Psi_{0}}\in \mathrm{Span}\{\ket{\bar{j}\bar{k}} : \bar{0}\leq \bar{j},\bar{k}\leq \overline{J-1}\}$. Then the action of the time evolution operator $U=SC$ is closed on the subspace $\mathrm{Span}\{\ket{\bar{j}\bar{k}} : \bar{0}\leq \bar{j},\bar{k}\leq \overline{J-1}\}$ of $\mathcal{H}_{DTQW}$ as follows:
\begin{align*}
U\ket{\bar{j}\bar{k}}
=
\begin{cases}
\displaystyle\frac{2}{d_{\bar{j}}}\left(\sum_{G_{\bar{j}}\sim G_{\bar{k^{\prime}}}}\sqrt{d_{\bar{j},\bar{k}}d_{\bar{j},\bar{k^{\prime}}}}\ket{\bar{k^{\prime}}\bar{j}}\right)
-\ket{\bar{k}\bar{j}}, &\text{if $\bar{j}\neq \bar{0}$ and $\bar{0}\leq \bar{k}\leq \overline{J-1}$},\\
-\ket{\bar{k}\bar{0}},&\text{if $\bar{j}=\bar{0}$ and $\bar{0}\leq \bar{k}\leq \overline{J-1}$}.
\end{cases}
\end{align*}
\end{thm}

\subsection{CTQW cases}
In this section, we consider the CTQW search on simple, connected graph $G$ with equitable partition $(G_{\bar{0}}, G_{\bar{1}}, \ldots ,G_{\overline{J-1}})$ with $n_{\bar{j}}=|V(G_{\bar{j}})|$ of $G$. We define the following uniform states related to the partition:
\begin{align*}
\ket{\bar{j}}=\frac{1}{\sqrt{n_{\bar{j}}}}\sum_{j\in V(G_{\bar{j}})}\ket{j},
\end{align*}
for $0\leq \bar{j}\leq J-1$. 
By direct calculation, we obtain
\begin{align*}
A_{G}\ket{\bar{j}}
&=
d_{\bar{j}\bar{j}}\ket{\bar{j}} + \sum_{G_{\bar{j}}\sim G_{\bar{k}}}d_{\bar{k}\bar{j}}\sqrt{\frac{n_{\bar{k}}}{n_{\bar{j}}}}\ket{\bar{k}}
=
d_{\bar{j}\bar{j}}\ket{\bar{j}} + \sum_{G_{\bar{j}}\sim G_{\bar{k}}}\sqrt{d_{\bar{k}\bar{j}}}\sqrt{\frac{n_{\bar{j}}d_{\bar{j}\bar{k}}}{n_{\bar{j}}}}\ket{\bar{k}}
\\
&=
d_{\bar{j}\bar{j}}\ket{\bar{j}} + \sum_{G_{\bar{j}}\sim G_{\bar{k}}}\sqrt{d_{\bar{j}\bar{k}}d_{\bar{k}\bar{j}}}\ket{\bar{k}},
\\
&=
\sum_{\bar{0}\leq \bar{k}\leq \overline{J-1}}\sqrt{d_{\bar{j}\bar{k}}d_{\bar{k}\bar{j}}}\ket{\bar{k}},
\end{align*}
where we use the relation $n_{\bar{j}}d_{\bar{j},\bar{k}}=n_{\bar{k}}d_{\bar{k},\bar{j}}$ coming from simplicity of the graph. 

Now we define a $J\times J$ matrix $\bar{A}_{G}$ as 
\begin{align*}
(\bar{A}_{G})_{\bar{j},\bar{k}} = 
\sqrt{d_{\bar{j}\bar{k}}d_{\bar{k}\bar{j}}}.
\end{align*}
Recall that the Hermitian matrix for DTQW search is defined by $H_{G}=|0\rangle \langle 0| + \gamma A_{G}$. If we define a $J\times J$ matrix $\bar{H}_{G}$ as 
\begin{align*}
\bar{H}_{G} = \mathrm{diag}(1, 0, \ldots , 0) + \gamma \bar{A}_{G},
\end{align*}
then the action of the Hermitian matrix $H_{G}$ is closed on the subspace $\mathrm{Span}\{\ket{\bar{j}} : \bar{0}\leq \bar{j} \leq \overline{J-1}\}$ as
\begin{align*}
H_{G}\ket{\bar{j}}
=
\sum_{\bar{0}\leq \bar{k}\leq \overline{J-1}}\left(\bar{H}_{G}\right)_{\bar{j},\bar{k}}\ket{\bar{k}},
\end{align*}
for $\bar{0}\leq \bar{j} \leq \overline{J-1}$. We also obtain $\ket{\Psi_{0}}=\frac{1}{\sqrt{N}}\sum_{\bar{j}=\bar{0}}^{\overline{J-1}}\sqrt{n_{\bar{j}}}|\bar{j}\rangle \in \mathrm{Span}\{\ket{\bar{j}} : \bar{0}\leq \bar{j} \leq \overline{J-1}\}$. This shows that
\begin{align*}
H(G)\ket{\Psi_{0}}
=
\bar{H}_{G}\left(\frac{1}{\sqrt{N}}\sum_{\bar{j}=\bar{0}}^{\overline{J-1}}\sqrt{n_{\bar{j}}}|\bar{j}\rangle \right),
\end{align*}
thus 
\begin{align*}
\exp\left(itH(G)\right)\ket{\Psi_{0}}
=
\exp\left(it\bar{H}_{G}\right)\left(\frac{1}{\sqrt{N}}\sum_{\bar{j}=\bar{0}}^{\overline{J-1}}\sqrt{n_{\bar{j}}}|\bar{j}\rangle \right).
\end{align*}
We summarize this fact as the following theorem:
\begin{thm}\label{thm_CTQW}
Consider the CTQW search on simple, connected graph $G$ with the equitable partition $(G_{\bar{0}}, G_{\bar{1}}, \ldots , G_{\overline{J-1}})$.  Then we have $\ket{\Psi_{0}}\in \mathrm{Span}\{\ket{\bar{j}} : \bar{0}\leq \bar{j}\leq \overline{J-1}\}$. Define a $J\times J$ matrix $\bar{H}_{G}$ as 
\begin{align*}
\bar{H}_{G}
=
\mathrm{diag}(1, 0, \ldots , 0) + \gamma \bar{A}_{G},
\end{align*}
where $\bar{A}_{G}$ is also $J\times J$ matrix with $(\bar{A}_{G})_{\bar{j},\bar{k}}= \sqrt{d_{\bar{j}\bar{k}}d_{\bar{k}\bar{j}}}$. Then the finding probability $\mathbb{P}_{N,t}(0)$ of the marked vertex $0\in V(G)$ at time $t$ is 
\begin{align*}
\mathbb{P}_{N,t}(0) = \left|\bra{\bar{0}}\exp\left(it\bar{H}_{G}\right)\left(\frac{1}{\sqrt{N}}\sum_{\bar{j}=\bar{0}}^{\overline{J-1}}\sqrt{n_{\bar{j}}}|\bar{j}\rangle \right)\right|^{2}.
\end{align*}
\end{thm}

By using Theorem \ref{thm_CTQW}, we obtain the following concrete example of CTQW search:
\begin{pro}\label{pro_CTQWsearch}
Let $G$ be the graph with marked vertex $0\in V(G)$ which is connected to all the vertices of  a $d$-regular graph with $N-1$ vertices then we obtain
\begin{align*}
\mathbb{P}_{N,t}(0)
=
\left( 1-\frac{1}{N} \right) - \left( 1-\frac{2}{N} \right) \cos^{2}\left( \frac{\sqrt{N-1}}{d}t \right),
\end{align*}
when we take $\gamma =1/d$. Particularly, if we take $T=(d\pi /2)/\sqrt{N-1}$ then we have $\mathbb{P}_{N,T}(0)=1-1/N$. The finding probability $\mathbb{P}_{N,t}(0)$ takes any values in $[1/N,1-1/N]$ with period $2T$.
\end{pro}
Proposition \ref{pro_CTQWsearch} shows that if we consider a graph with $d=\mathcal{O}(N)$ such that the complete graph $K_{N}$ then $T=\mathcal{O}(\sqrt{N})$ with $\gamma =\mathcal{O}(1/N)$. In this case, the reduced matrix $\bar{H}_{G}$ is a $2\times 2$ matrix (simplest case). Finding concrete examples which reduced matrix $\bar{H}_{G}$ is a general $J\times J$ matrix and the search is successful might be an interesting future problem. 

{\bf Proof of Proposition \ref{pro_CTQWsearch}.}

In this case, we can consider an equitable partition $(G_{0}, G_{1})$ with $V(G_{0})=\{0\}, V(G_{1})=\{1,\ldots ,N-1\}$ and $d_{0,0}=0, d_{0,1}=N-1, d_{1,0}=1, d_{1,1}=d$. By Theorem \ref{thm_CTQW} we have the reduced matrix
\begin{align*}
\bar{H}_{G}
=
\begin{bmatrix}
1 & \gamma \sqrt{N-1} \\
\gamma \sqrt{N-1} & \gamma d
\end{bmatrix}.
\end{align*}
In order to obtain eigenvalues and eigenvectors of $\bar{H}_{G}$, we deal with the following equation:
\begin{align*}
\bar{H}_{G}\left( \alpha |\bar{0}\rangle + \beta |\bar{1}\rangle \right) = \lambda \left( \alpha |\bar{0}\rangle + \beta |\bar{1}\rangle \right) .
\end{align*}
This equation is equivalent to
\begin{align*}
\begin{bmatrix}
1 & \gamma \sqrt{N-1} \\
\gamma \sqrt{N-1} & \gamma d
\end{bmatrix}
\begin{bmatrix}
\alpha \\
\beta
\end{bmatrix}
=
\lambda 
\begin{bmatrix}
\alpha \\
\beta
\end{bmatrix}.
\end{align*}
If we set $\gamma =1/d$ then the eigenvalues and eigenvectors $\bar{H}_{G}\ket{v_{\pm}}=\lambda_{\pm}\ket{v_{\pm}}$ are obtained by 
\begin{align*}
\lambda_{\pm}
&=
1\pm \frac{\sqrt{N-1}}{d},
\\
\ket{v_{\pm}}
&=
\frac{\ket{\bar{0}}\pm \ket{\bar{1}}}{\sqrt{2}}.
\end{align*}
Therefore we obtain spectral decomposition
\begin{align*}
\exp\left(it\bar{H}_{G}\right)
=
e^{it\lambda _{+}}| v_{+} \rangle \langle v_{+} |
+
e^{it\lambda _{-}}| v_{-} \rangle \langle v_{-} |.
\end{align*}
On the other hand,
\begin{align*}
&\bra{v_{\pm}}\left\{\frac{1}{\sqrt{N}}\left(\ket{\bar{0}}+\sqrt{N-1}\ket{\bar{1}}\right)\right\}
=
\frac{1}{\sqrt{2N}}\left(1\pm \sqrt{N-1}\right),
\\
&\langle \bar{0}\ket{v_{\pm}}
=
\frac{1}{\sqrt{2}}.
\end{align*}
Then we have
\begin{align*}
&
\langle \bar{0}|\exp\left(it\bar{H}_{G}\right)\left\{\frac{1}{\sqrt{N}}\left(\ket{\bar{0}}+\sqrt{N-1}\ket{\bar{1}}\right)\right\}
\\
&=
e^{it}\left\{ \frac{1}{\sqrt{N}}\cos\left( \frac{\sqrt{N-1}}{d}t \right) + i\sqrt{\frac{N-1}{N}}\sin\left( \frac{\sqrt{N-1}}{d}t \right) \right\}.
\end{align*}
As a consequence, we obtain
\begin{align*}
\left|\langle \bar{0}|\exp\left(it\bar{H}_{G}\right)\left\{\frac{1}{\sqrt{N}}\left(\ket{\bar{0}}+\sqrt{N-1}\ket{\bar{1}}\right)\right\}\right|^{2}
=
\left( 1-\frac{1}{N} \right) - \left( 1-\frac{2}{N} \right) \cos^{2}\left( \frac{\sqrt{N-1}}{d}t \right).
\end{align*}
\qed

\section{Summary}
In this paper, we show reduction methods for search algorithms on graphs using QWs. By using the equitable partition for the graph, we determine ``effective subspace'' for the search algorithm to reduce the size of the problem in both DTQW search and CTQW search. It can be an interesting future problem that determining conditions of equitable partition which induce successful DTQW search and CTQW search.

\par
\
\par\noindent
{\bf Acknowledgments.} 

The author was supported by the Grant-in-Aid for Young Scientists (B) of Japan Society for the Promotion of Science (Grant No. 16K17652). 

\begin{small}

\end{small}

\end{document}